\documentclass[10pt,letterpaper]{article}
\usepackage{opex3}
\usepackage{cite}
\usepackage{multirow}

\begin{document}

%%%%%%%%%%%%%%%%%% title page information %%%%%%%%%%%%%%%%%%
\title{Photoluminescence in Ga/Bi co-doped silica glass}

\author{Igor Razdobreev,$^{1}$ Hicham El Hamzaoui,$^{2}$  \mbox{Vladimir B. Arion,$^{3}$}  and Mohamed Bouazaoui$^2$}

\address{
$^1$CERLA, PHLAM UMR CNRS 8523 University Lille-1, 59655 Villeneuve d'Ascq, France\\
$^2$IRCICA - UMR8523/FR3024 CNRS, Parc de la Haute Borne, 50 av. Halley, Villeneuve d'Ascq, 59658, France\\
$^3$Institute of Inorganic Chemistry, University of Vienna, Waehringer Str. 42, Vienna, A-1090, Austria\\
}

\email{Igor.Razdobreev@univ-lille1.fr} %% email address is required


\begin{abstract}
Bismuth-Gallium co-doped silica glass fiber preform was prepared from nano-porous silica xerogels using
a conventional solution doping technique with a heterotrinuclear complex and subsequent sintering. Ga-connected
optical Bismuth active center (BAC) was identified as the analogue of Al-connected BAC. Visible and infrared photoluminescence (PL)
were investigated in a wide temperature range of 1.46\,-\,300\,K. Based on the results of the continuous wave (CW) and time resolved (TR)
spectroscopy we identify the centers emitting in the spectral region of 480\,-\,820\,nm as Bi$^+$ ions. The near infrared (NIR) PL 
around 1100\,nm consists of two bands. While the first one can be ascribed to the transition in Bi$^+$ ion, the second band is
presumably associated to defects.
We put in evidence  the energy transfer (ET) between Bi$^+$ ions and the second NIR emitting center via quadrupole-quadrupole
and dipole-quadrupole mechanisms of interactions. Finally, we propose the energy level diagram of Bi$^+$ ion interacting
with this defect.
\end{abstract}


%%%%%%%%%%%%%%%%%%%%%%% References %%%%%%%%%%%%%%%%%%%%%%%%%

%%%%%%%%%%%%%%%%%%%%%%%%%%  body  %%%%%%%%%%%%%%%%%%%%%%%%%%
\section{Introduction}
Over the past decade there has been increasing interest in the development of Bi-doped fiber lasers (BFL) and amplifiers (BFA).
Notwithstanding the significant progress achieved in recent years \cite{Bufetov09}, such devices suffer from a number of drawbacks. 
The very low levels of Bismuth doping and, as a consequence, significant fiber length (typically 80\,-\,100\,m) are necessary to ensure
the efficient BFL and BFA operation. It is generally believed that the poor understanding of the nature of the
PL in Bi-doped glasses is the main reason restraining the development of the efficient devices. Indeed, since the first demonstration
of the NIR PL in Bi-doped silica glasses \cite{Murata99,Fujimoto01} and up to
now there is no consensus  on the nature of the PL in Bi-doped glasses (see, for instance the reviews on the subject
 \cite{Peng11,Dianov12} and references therein).
In this context the search and investigation of the new glass compositions provide a significant insight into the
understanding of the nature of luminescent centers in Bi-doped silica glasses. This is even more important than the attempts
of the empirical optimization of the BFL and BFA efficiency without the deep understanding of the PL nature in Bi-doped glasses.

In the present work we investigate the luminescent properties of Ga/Bi co-doped silica in a wide temperature range. We show 
that the PL centers in this material are the analogues of those in Al/Bi co-doped silica glass. We also put in evidence that when
excited in the visible, the NIR PL centers emitting in the range of 1000\,-\,1200\,nm are mainly populated via the efficient
ET from the excited states of the centers absorbing and emitting in the visible part of the complex PL spectrum. While the latter
centers can be identified as Bi$^+$ ions, the NIR PL, most probably, should be assigned to the defects, as it was suggested earlier 
\cite{Sharonov08,Sharonov09}.  Based on our experiments, we propose a consistent energy level diagram, which explains the
experimental results in a weakly doped with Bismuth and Ga(Al) silica glasses. 

We would like to emphasize, that the present work gives an overview of many experiments that cannot be described in detail in a single article. 
The detailed description and analysis of the temperature dependence of the decay kinetics at 737\,nm, temperature dependence of the
anti-Stokes PL and the detailed investigation of the zero-phonon line observed at low temperature remain out of the scope of the present article
and will be published elsewhere. 

\section{Experimental}
\subsection{Sample preparation and characterization}
The novel bulk Ga/Bi co-doped pure silica glass was produced from nano-porous silica xerogel doped with a  synthesized
heterotrinuclear molecular precursor, namely, [Bi$_2$(HSal)$_6$Ga(acac)$_3$], where Hsal = O$_2$CC$_6$H$_4$-2-OH and Hacac = CH$_3$COCH$_2$COCH$_3$.
The step of the high temperature drawing was omitted in the present experiments and the samples with
the dimensions of 2\,x\,3\,x\,5\,mm$^3$ were cut, polished and analyzed after the step of the sintering at 1300\,$^{\circ}\mathrm{C}$.
The content of Bismuth and Gallium was measured as 35\,ppm and 138\,ppm, respectively. The details of the fabrication technology and
the content measuring procedure can be found in \cite{HEH10,HEH11,R12}. 

\subsection{Experimental set-up}
The details of our experimental PL set-up were reported in \cite{R12}. The present experiments were performed with the closed cycle
helium cryostat (SpectromagPT, Oxford Instr.) operating in a wide temperature range of 1.4\,-\,300\,K. The thermal stability of the
samples attached to the holder of the variable temperature insert  was about of 0.01\,K except the range from 4.2 to 10\,K where 
the thermal stability was $\sim$\,0.05\,K. In CW PL experiments the following laser sources were employed: Verdi (6 W, 532 nm, 
Coherent Inc.), tunable CW Ti:Sapphire (685\,-\,1000\,nm, Model 899, Coherent Inc.) and tunable Yb-doped fiber laser (custom laser,
1060\,-\,1100\,nm, up to 10\,W, Manlight, France). The additional pulsed laser source was the optical parametric oscillator (OPO) Opolette
355 LD (8\,ns, Opotek Inc.). In the experiments the pulse energy from the OPO was maintained constant, about of 100 $\mu$J per pulse.
In the visible range of the spectrum the PL decay kinetics were measured with the high speed photomultiplier with the transit time spread
(TTS) as low as 120\,ps and free from the afterpulsing (PMC-100-20, Becker \& Hickl GmbH). The anti-Stokes spectra were acquired with
the help of the cell FTIR600 (Linkam Scientific Instr.) with the temperature stability better than 0.1\,K.
The spectral and time resolution were variable from 0.7 to 3\,nm and from 1\,ns to 1.024\,$\mu$s, respectively, depending on the spectral
range and the operation mode (CW or TR).  

\section{Results and discussion}
\subsection{CW experiments}
\begin{figure}[htb]
\centerline{\includegraphics[width=12cm]{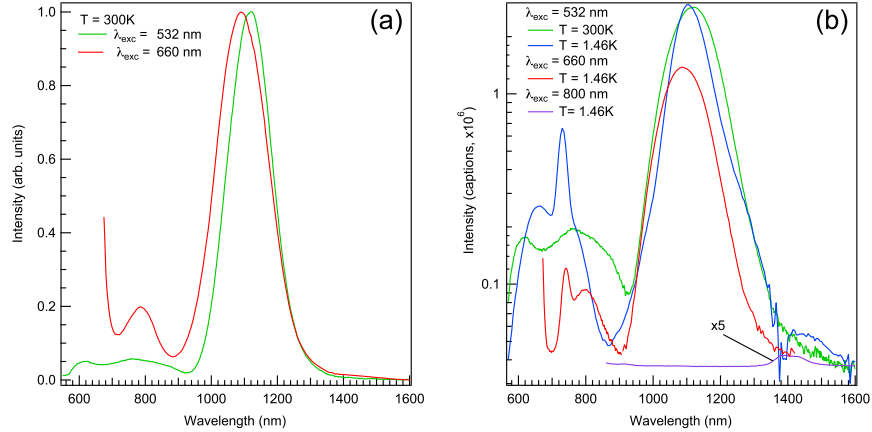}}
\caption{CW PL spectra. a) Normalized room temperature PL spectra under excitation at 532 and 660 nm;
b) Comparison of the PL spectra at $\lambda_{exc}$\,=\,532\,nm (T\,=\,300 and 1.46\,K), $\lambda_{exc}$\,=\,660\ nm (T\,=\,1.46\,K)
and $\lambda_{exc}$\,=\,800\ nm (T\,=\,1.46\,K).}
\end{figure}
In Fig.\,1(a) we show the normalized room temperature continuous wave (CW) spectra in logarithmic scale under excitation at 532 and
660\,nm. At room temperature the excitation at 532\,nm reveals the band at 615\,nm, which was always assigned to
Bi$^{2+}$ ions \cite{Dianov12}, the complex band with the maximum at 756\,nm and the very intense NIR band with the maximum at 1140\,nm and
the full width at half maximum (FWHM) about of 160\,nm (hereafter this band will be referenced as P0 band). The excitation at 660\,nm
reveals only two PL bands at 787 and 1090\,nm, the latter with FWHM about of 185\,nm. Note the considerable blue shift (50\,nm) and
the increase of the width of P0 band comparing to that under excitation at 532\,nm.  In Fig.\,1(b) we show in a logarithmic scale \lq\lq as is" spectra under
excitation at 532\,nm but recorded at two temperatures, 300 and 1.46\,K and the spectra under excitation at 660 and 800\,nm (1.46\,K). First, 
the excitation at 800\,nm reveals the weakly pronounced maximum at 1400\,nm due to the centers in a silica sub-lattice \cite{R10}. 
This difference in comparison to the recently reported by us results on the PL in Al/Bi co-doped silica glass \cite{R12},  where only
Al-conected centers were observed, can be attributed to the difference in the Al/Bi and Ga/Bi doping ratio, 80 and 4, 
respectively. Most probably, the low chemical reactivity of Ga with the host silica, in comparison to Al, leads to its enhanced volatility
and, as a consequence, to its lower content in the sintered glass. The latter results in the increased probability of the formation of the
NIR centers in a silica sub-lattice.  Nevertheless, the content of the Bismuth related centers in a silica sub-lattice remains extremely low and 
they are not considered in the present paper. Second, the band P0 is absent under excitation at 800\,nm. Third, it is clearly seen that
the band at 615\,nm completely disappears at low temperature under excitation at 532\,nm, while the complex band around of 740\,nm
reveals its triplet structure. The triplet structure of this band was pointed out previously in \cite{R11}, 
where Bi-doped germano-alumino-phospho-silicate fiber preform was investigated at low temperature.  

\begin{figure}[htb]
\centerline{\includegraphics[width=12cm]{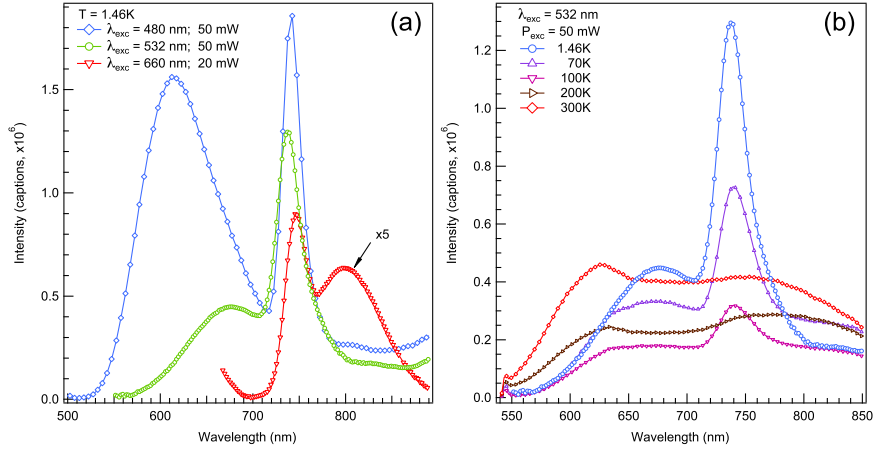}}
\caption{CW PL spectra in the spectral range of 500\,-\,900\,nm. a) T\,=\,1.46\,K; $\lambda_{exc}$\,=\,480, 532 and 660\,nm.
b) $\lambda_{exc}$\,=\,532 nm, variable temperature.}
\end{figure}
The temperature dependence of the PL in the wavelength region of 500\,-\,900\,nm was investigated in more details.
Fig.\,2(a) shows the PL spectra recorded at 1.46\,K at different excitation wavelengths. Though we do not provide the detailed
deconvolution, it is seen from Fig.\,2(a) that both red bands, peaked at 615 and 670\,nm, are present in the spectrum when the
sample is excited at 480\,nm. Under excitation at 532\,nm the  band at 615\,nm is absent, as it was mentioned above. Fig.\,2(b) shows
the temperature dependence of the spectrum under excitation at 532\,nm. At 300\,K the band at 615\,nm is dominant in the CW spectrum. 
The decrease of the temperature results in the decrease of its relative intensity so that below 70\,K this band completely disappears. The same
behavior was observed in the temperature dependent time resolved spectra (see the next sub-section). 
\begin{figure}[htb]
\centerline{\includegraphics[width=12cm]{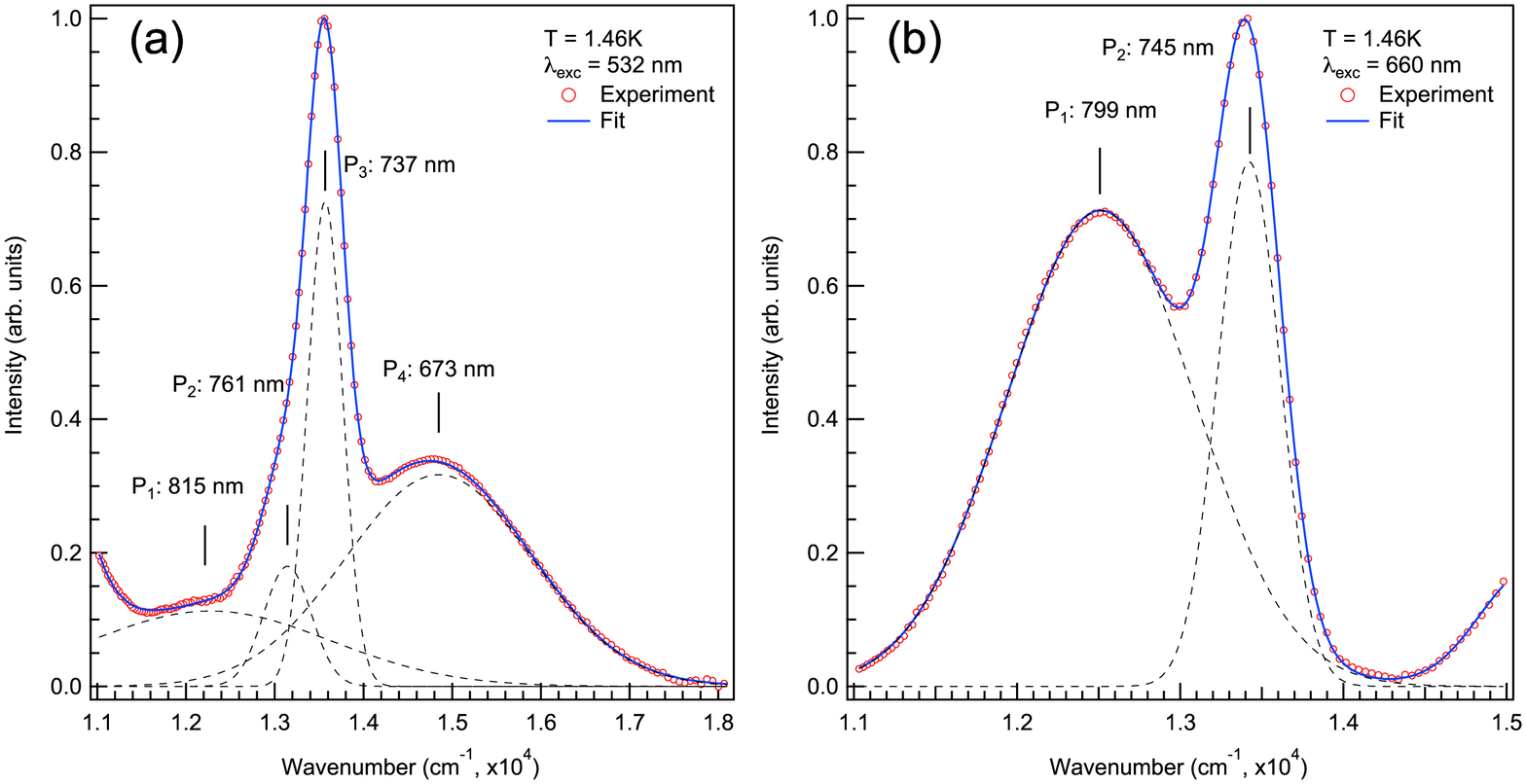}}
\caption{Gaussian decomposition of the PL bands in the region of 550\,-\,900\,nm
at T\,=\,1.46K. a) $\lambda{_{exc}}$\,=\,532\,nm; b) $\lambda{_{exc}}$\,=\,660\,nm.
Spectral resolution 0.8\,nm.}
\end{figure}

In Fig.\,3(a) and 3(b) we show the Gaussian deconvolution of the spectrum recorded at 1.46\,K under excitation at 532 and
660\,nm, respectively. In this decomposition the dummy bands (not shown in the graphs) were introduced to model the short-wavelength
shoulder of  the main NIR PL band (for 532\,nm excitation) and the diffuse scattering from the pump (for 660\,nm excitation). For the sake 
of completeness all the peak parameters are collected in Table\,1. One can see that there are \textit{four} bands in the PL
spectrum in the range of 550\,-\,900\,nm under excitation at 532\,nm at 1.46\,K, while the excitation at 660\,nm, results in only \textit{two}
PL bands and not in three as it could be expected. Because the position and intensity of all PL bands depend strongly on the excitation
wavelength, this CW experiment does not address exactly the question \lq\lq which PL band disappears in the spectrum under excitation at
660\,nm in comparison to the spectrum excited at 532\,nm?", though, most probably, it should be the PL band P3. This assumption was
confirmed by the time resolved experiments (see the next subsection).   

\begin{figure}[htb]
\centerline{\includegraphics[width=12cm]{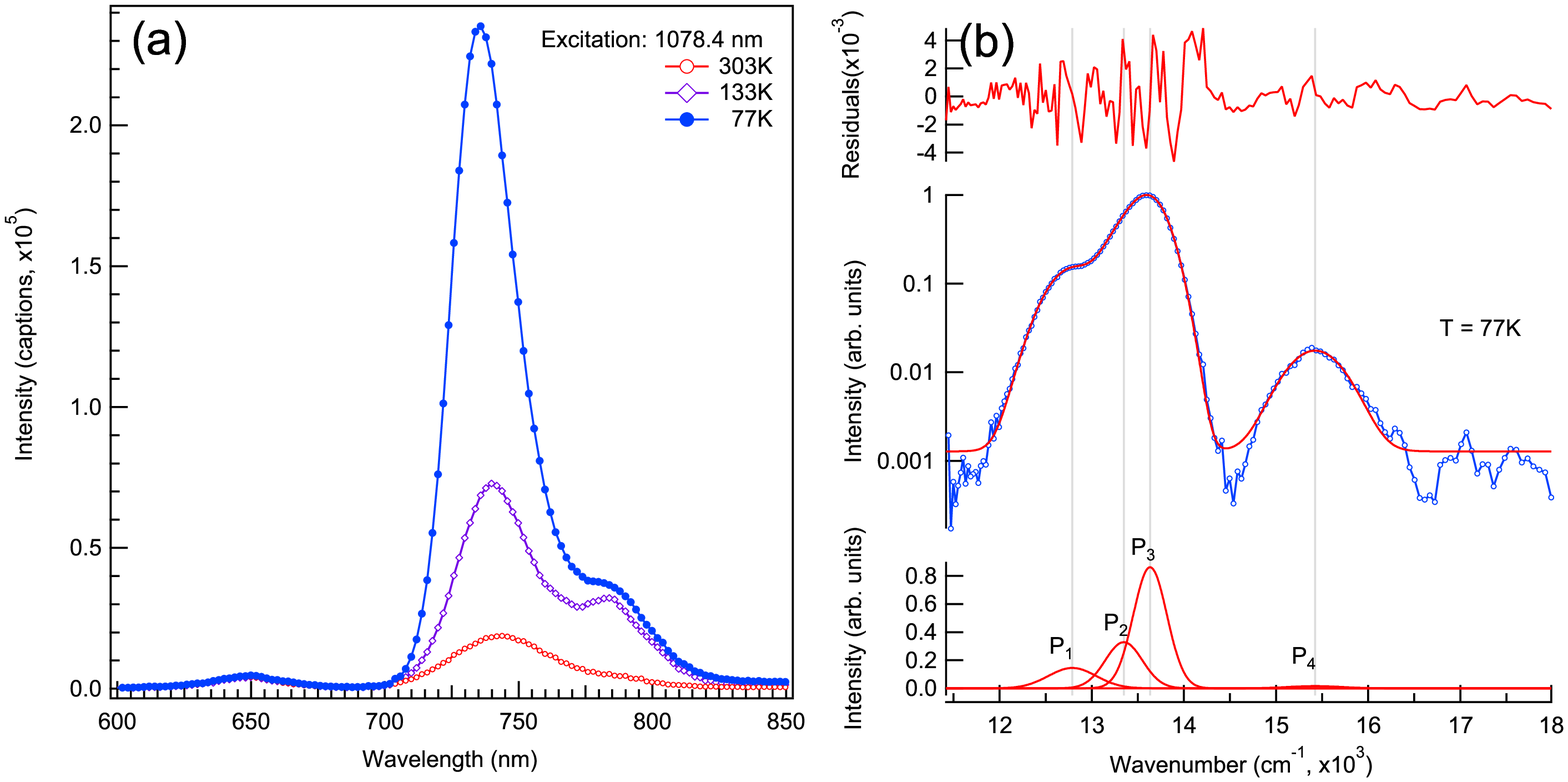}}
\caption{Anti-Stokes PL in the region of 600\,-\,850\,nm, $\lambda{_{exc}}$\,=\,1078.4\,nm,
P${_{exc}}$\,=\, 1\,W, spectral resolution 1\,nm.
a) various temperatures, \lq\lq as is" spectra ; b) Gaussian decomposition of the spectrum recorded at T\,=\,77\,K.
The spectrum was corrected and normalized (see the data in Table\,1).}
\end{figure}

All the bands observed under excitation at 532\,nm at low temperature can be readily observed in the anti-Stokes
spectrum at the excitation WL's in the range of 1060\,-\,1100\,nm. The examples of such spectra are shown in Fig.\,4(a)
for the excitation WL 1078.4\,nm for different temperatures. In Fig.\,4(b) we show the Gaussian deconvolution of the
spectrum recorded at 77\,K, the details of the band parameters are also reported in Table\,1. We would like to point out the following
features: (i) the intensity of the PL bands P1\,-\,P3 in the anti-Stokes spectrum increases with lowering the temperature and
(ii) the temperature independent intensity of the band P4. Such a behavior was observed down to 1.46\,K and it correlates well with the
temperature dependence of the lifetimes of the bands P3 and P4 that are discussed in the next subsection.  

\begin {table}[h]
\caption {Summary of PL bands observed at 1.46\,K under excitation at 532 and 660\,nm and in the anti-stokes spectrum at 77\,K} \label{Tab1} 
\begin{center}
\begin{tabular}{ |c|c|c|c|c|}
\hline
Excitation WL  & Peak no. & Location  & Height  & Width  \\
(nm) & & (cm$^{-1}$/nm) & (arb. units) & (cm$^{-1}$) \\ \hline
\multirow{4}{*}{532} & P1 & 12276 / 815 & 0.113 & 1916 \\
&  P2 & 13144 / 761 & 0.18 & 379 \\
& P3 & 13568 / 737 & 0.726 & 281 \\
& P4 & 14849 / 673 & 0.317 & 1488 \\ \hline
\multirow{2}{*}{660} & P1 & 12512 / 799 & 0.712 & 816 \\
 & P2 & 13422 / 745 & 0.786 & 270\\ \hline
 \multirow{4}{*}{1078.4 (anti-Stokes)} & P1 & 12789 / 782 & 0.145 & 367 \\
  &  P2 & 13348 / 749 & 0.328 & 285 \\
 & P3 & 13633 / 734 & 0.862 & 258 \\
 & P4 & 15424 / 648 & 0.016 & 414 \\ 
\hline
\end{tabular}
\end{center}
\end {table}

\begin{figure}[h]
\centerline{\includegraphics[width=12cm]{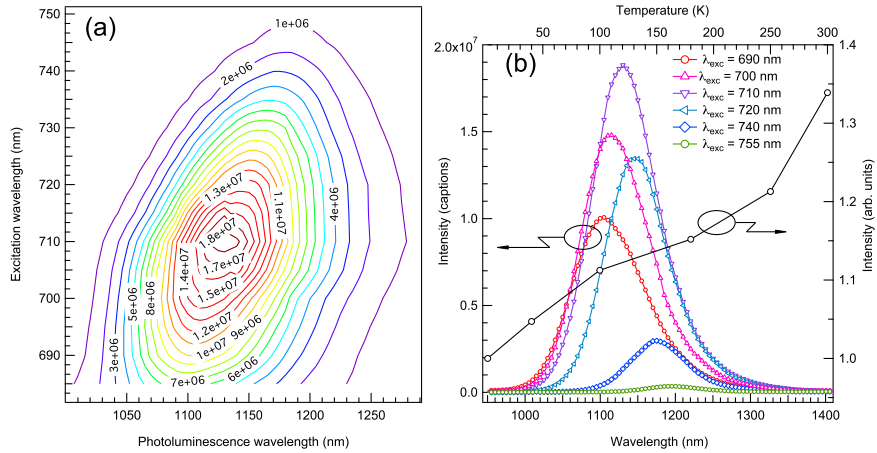}}
\caption{a) Contour plot of the PL intensity in Ga/Bi co-doped silica glass, T = 1.46\,K; 
b) In color: spectra at the selected excitation WL's, T = 1.46\,K. Black circles: normalized temperature dependence
of the integral intensity under excitation at 710\,nm. The integral intensities were normalized to that at 1.46\,K, the line 
is drawn to guide the eyes). }
\end{figure}

In \cite{Dvoyrin08} Dvoyrin \textit{et al.} discussed the possible origin of the low laser efficiency in aluminosilicate fibers operating in the
wavelength range of  1150\,-\,1250\,nm. They assumed, that the corresponding excited state level is constituted of two sub-levels with the
crossover between two adiabatic potential-energy surfaces (APES's). According to this assumption, the rise of the temperature leads to the
depletion of the low (lasing) sub-level population that in turn leads to the reduced efficiency of the fiber laser, as it was observed in the
laser experiments. On the other hand, in \cite{Bulatov10} the same group discussed the opposite point of view, which
consists in introducing of two PL centers in different environments, namely, the Bi-ion in tetrahedral and octahedral environments. Then, to explain the
low  laser efficiency at room temperature, it is possible to assume the resonance or phonon assisted ET between two centers.
It is worth noting that, to the best of knowledge, the ET or up-conversion was not considered as a possible source of the low efficiency
of fiber lasers up to now.  We performed the detailed investigation of the NIR PL band under various excitation WL's in the range of
685\,-\,760\,nm  at 1.46\,K (the short-wave excitation edge was limited by the Ti:Sapphire tuning range). In this experiment the excitation power 
was maintained at the constant level of 50 mW at all WL's, with the exception of 685\,nm. At this WL the laser delivered only 10\,mW, so the 
data obtained at this excitation WL were simply re-normalized. The results are shown in Fig.\,5(a) and Fig.\,5(b). It is seen that the contour plot,
corresponding to the three-dimensional  excitation-PL spectrum is smooth and it does not reveal any indication on the presence
of two distinct emitting centers. The maximum of NIR PL  is reached under excitation at 710\,nm. At this excitation WL the maximum of the band P0 
and its  FWHM are 1130 and 105\,nm, respectively.  Additionally, we measured the temperature dependence of the integral intensity of this NIR PL band
under excitation at fixed WL 710\,nm, which is also shown in Fig.\,4(b). It is seen that the integral intensity monotonously rises with
the temperature. As it will be shown in the next subsection, this behavior is a direct consequence of the ET from the excited
state corresponding to the PL band P3 to another center. 

\begin{figure}[htb]
\centerline{\includegraphics[width=8cm]{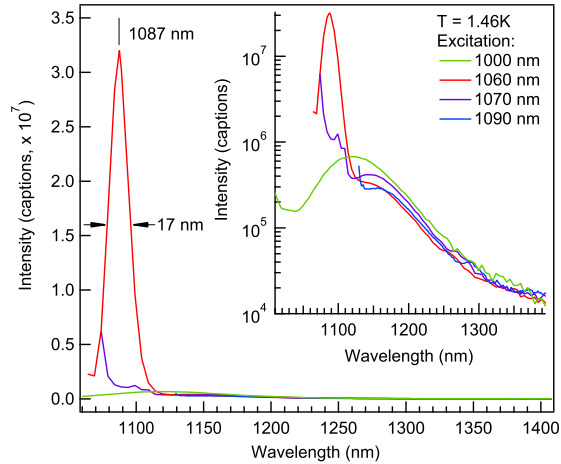}}
\caption{PL spectra recorded under excitation at 1000, 1060, 1070 and 1090\,nm. Excitation power 50\,mW, 
T\,=\,1.46\,K, spectral resolution 1.5\,nm.
Main window: \lq\lq as is" spectra, linear scale. Inset:  the same spectra in a semi-log scale.}
\end{figure}

The structure of P0 band can be revealed at low temperature by scanning
the excitation WL from 1000 to 1100\,nm. The corresponding spectra are shown in Fig.\,6. It is seen, that the intense PL
band with the maximum at 1087\,nm and FWHM about of 17\,nm ($\sim$\,143\,cm$^{-1}$) appears at the excitation WL 1060\,nm. 
Its intensity rapidly decreases with the small detuning ($\pm$10\,nm) from this excitation WL, so that only a wide and low intensity
band can be observed. The maximum of this low intensity band depends on the excitation WL, but its intensity remains nearly constant, 
as it is seen in the inset of Fig.\,6. We believe, that the strong intense band is the inhomogeneously broadened zero-phonon line (ZPL)
corresponding to the lowest excited state of Bi$^+$ ion, while the wide and low intensity band corresponds to the defects associated
to Bismuth doping. The high intensity of ZPL in comparison to the sideband and its small shift from the excitation WL
($\sim$\,234\,cm$^{-1}$ and 1060\,nm, respectively) mean that the Pekar-Huang-Rhys parameter \textit{S} does not exceed 0.5
taking into account that the maximum of the phonon density states in a pure sol-gel silica glass was found at 430\,cm$^{-1}$ \cite{HEH10}.  
The rapid attenuation of the intensity of ZPL with the detuning of the excitation WL from 1060\,nm indicates that the corresponding APES 
is very shallow.

\subsection{Time resolved experiments}

Measurements of the decay kinetics of the PL bands were performed in the temperature range of 1.46\,-\,300\,K at various excitation WL's.
In Fig.\,7 we show the time resolved spectra recorded at 300 and 1.46\,K (Fig.\,7(a) and 7(b), respectively) in the range of 550\,-\,900\,nm
under excitation at 532\,nm. The comparison with corresponding CW spectra recorded at various temperatures (Fig.\,2(b)) shows that the
short WL band with the maximum around of 615\,nm can be clearly identified at room temperature in the time resolved spectrum. But at low
temperature, as in the CW spectrum, this band completely disappears. Usually, this red band which appears in various Bi-doped glasses in the range
of 600\,-\,630\,nm is assigned to Bi$^{2+}$ ions (see, for instance \cite{Peng11}), but the complete disappearance of this band at low temperature
cannot be explained in this assumption. Such a temperature behavior allows us to doubt the correctness of the assignment of this band to the
transition $^2$P$_{3/2}$\,$\rightarrow$\,$^2$P$_{1/2}$ in Bi$^{2+}$ ions \cite{Blasse94} as it has been made in all previous studies
and it seems to be accepted as a proven fact \cite{Zhou08}. This behavior can be understood in the assumption of the crossover of two
APES of the same center or in the assumption of the double-well potential-energy surface. Finally,
we note that the narrow band with the maximum at 737\,nm (P3 according to Fig.\,3(a) and Table\,1) exhibits a very long lifetime at low
temperature in comparison to that at 300\,K. Below we will pay a special attention to the decay kinetics of the band P3.

\begin{figure}[htb]
\centerline{\includegraphics[width=11.5cm]{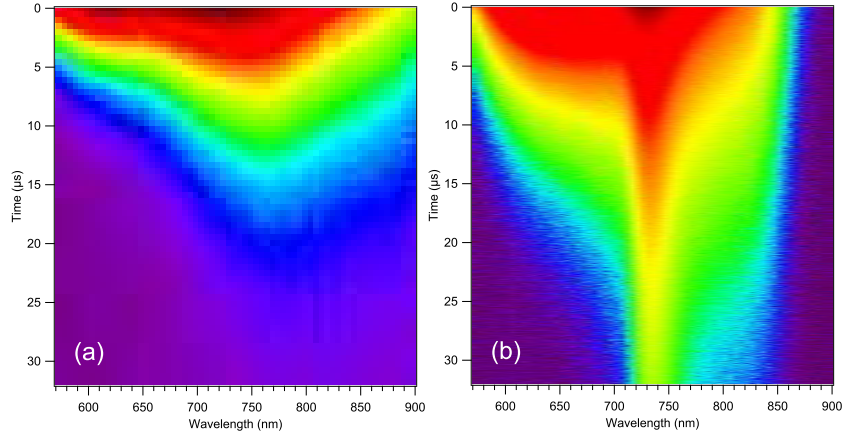}}
\caption{Time resolved PL spectra in the WL range 550\,-\,900\,nm, $\lambda{_{exc}}$\,=\,532\,nm, spectral resolution 1\,nm, time resolution 32\,ns. (a) T\,=\,300\,K;
(b) T\,=\,1.46\,K.}
\end{figure}

The comparison of the decay kinetics for the selected recording WL's under excitation at 532\,nm and at different temperatures are shown
in Fig.\,8. It is seen that at any temperature  all the decays exhibit a non-exponential behavior. For the recording WL's 670 and 780\,nm
(Fig.\,8(a) and Fig.\,8(b), respectively) the bi-exponential fits at 1.46 and 300\,K were given to illustrate (i) the non-exponential character of
the decays, and (ii) the temperature dependence of the lifetimes, which is very week in comparison to that for the P3 band at 737\,nm (Fig.\,7(c)).
The latter exhibits a very strong temperature dependence. While at high temperatures (200\,-\,300\,K) the decay of this band can be
approximated by the bi-exponential decay, at low temperature this approach is simply impossible (note the logarithmic scale of the vertical
axis). The strong temperature dependence of the decay kinetics can be also connected to the crossover of corresponding to this
transition APES with another one. This aspect requires a special consideration which is out of the scope of the present article and will be addressed elsewhere. 
Here we provide the detailed analysis of the decay kinetics at the lowest temperature, 1.46\,K.

\begin{figure}[h]
\centerline{\includegraphics[width=12cm]{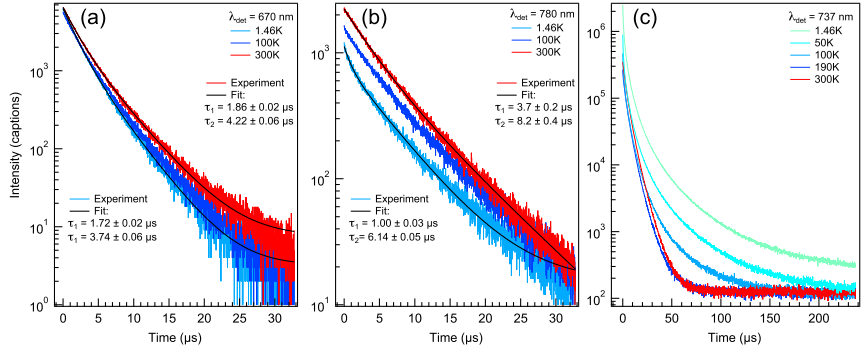}}
\caption{PL decay kinetics at the selected WL's and various temperatures. $\lambda{_{exc}}$\,=\,532\,nm, spectral resolution 1\,nm, time resolution 32\,ns. 
(a) 670\,nm; (b) 780\,nm; c) 737\,nm.}
\end{figure}

It should be emphasized that the analysis of the complex decay kinetics, like that shown in Fig.\,8(c) at 1.46\,K, for example, is the very
difficult task because  the exact nature of the emitting center and its environment are unknown \textit{a priori}. Sometimes the experimentalists
use the stretched exponential function \cite{Johnston06} which can be put in various analytical forms. This type of kinetics arises when
there exists some distribution in the lifetimes of the relaxing species. Though Hughes \textit{et al.} \cite{Hughes08} used the stretched exponential
function to fit the PL decay kinetics in Bi-doped lead-germanate glasses, the justification of this approach to Bi-doped glasses remains, in
our opinion, an open question. On the other hand, Romanov \textit{et al.} \cite{Romanov11} have recently shown that the first stage of the PL
kinetics in Bi-doped fluoride glass obeys the F\"{o}rster law which implies that the main relaxation channel of the excited species 
is the ET to some quenchers via dipole-dipole interaction \cite{Forster}. It is worth noting here that in both works the heavy doped samples
were investigated and the decay kinetics were measured in the NIR PL spectral range and at room temperature only in \cite{Hughes08} or
at the lowest temperature 77\,K in \cite{Romanov11}. Also, the strong temperature dependence of the decay forms was observed in the
latter work (see Fig.\,5 in the cited article) which is similar to the behavior shown in Fig.\,8(c) of the present work.
Recently, the pure silica glass prepared from nano-porous silica xerogels was also doped with Er$^{3+}$\cite{HEH11} and
Yb$^{3+}$\cite{Baz13}. In both cases the measured decay kinetics exhibited a single exponential decay even at the relatively high doping
levels. In the present study the doping levels is significantly lower, furthermore, the P3  band appears as a very narrow (see Table\,1).
For these reasons  we believe that the fit of the decay kinetics of the PL band P3 to the
stretched exponential function is most probably unjustified and we analyze this kinetics in the assumption of the ET phenomena between
donors and acceptors. 

\begin{figure}[htb]
\centerline{\includegraphics[width=12cm]{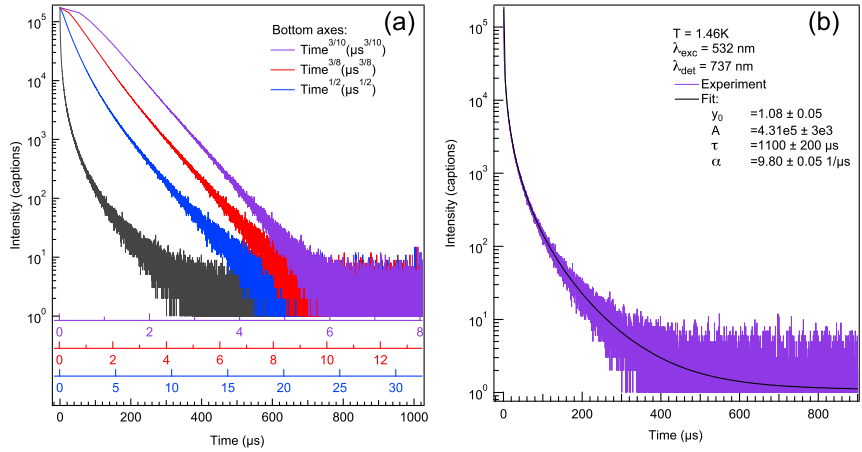}}
\caption{Decay kinetics of the band P3 (737\,nm). $\lambda{_{exc}}$\,=\,532\,nm, spectral resolution 1\,nm, time resolution 32\,ns. a) Normalized plots 
as a function of $t^{3/S})$  (the color of the particular plot corresponds to the color of the bottom axis); b) Experimental decay kinetics and the fit in the assumption
of the quadrupole-quadrupole nature of the donor-acceptor interaction.}
\end{figure}

It is well known \cite{Watts,Burshtein} that the decay kinetics in the presence of ET can be put in the form: $I=I_0\exp(-t/\tau_{0}-(\alpha t)^{d/S})$, where 
$\tau_{0}$ is the intrinsic lifetime of the excited state of the donor, $\alpha$ is the parameter related to the set of the microparameters; d = 3 for the three-dimensional
system and S = 6, 8, 10 is the multiplicity of the dipole-dipole, dipole-quadrupole and quadrupole-quadrupole interactions between the donor and acceptor. This equation is valid only at low
concentrations of donors, which is the case of the studied here samples. At high donor concentration the donor-donor interaction becomes important so that the migration of the excitation appears 
in the system of donors which can change the form of the law \cite{Sakun72}. The examples of the analysis of the experimental decay kinetics in rare-earth doped crystals
and glasses can be found in \cite{Voron'ko76,Avanesov83}. As it follows from the above equation, in the assumption of the very long intrinsic lifetime of the excited state of the donor
(the transition is forbidden) the normalized decay kinetics plotted as a function of $t^{3/S}$ in logarithmic scale, should be approximately linear for the corresponding type of
the interaction between donor and acceptor. These dependencies for the transition at 737\,nm are shown in Fig.\,9(a). It is seen that with the exception of the short initial
time range ($\sim$\,5\,$\mu$s) the logarithmic plot of the normalized decay as a function of $t^{3/10}$ is indeed very close to linear. In Fig.\,9(b) we show the result 
of the numerical fit to the above function taking into account the term corresponding to the intrinsic lifetime and the term corresponding to the quadrupole-quadrupole 
nature of the interaction between donor and acceptor, the numerical values are shown in the Figure. The fit was performed with a standard non-linear Levenberg-Marquardt
method with weights corresponding to Poisson photon-counting statistics \cite{Press} and taking the 5\,$\mu$s offset from the first (t\,=\,0) channel. There are two reasons we made
such an ofset: first, as it can bee seen from the Fig.\,3(a), there is an important contribution of the short-lived PL bands P1, P2 and P4 to the PL at 737\,nm; the second, and more important one,
is the relatively high initial intensity of P3 band, which leads to the partial overload of the photomultiplier. Though the reduction of the photon flux can be easily made, it leads to the lowering of
the signal to noise ratio, especially at the long record times. It is clearly seen, that the fit gives the excellent approximation to the experimental data. Note also the high dynamic range
(more then 5 orders of magnitude) and the very long record time.

\begin{figure}[htb]
\centerline{\includegraphics[width=12cm]{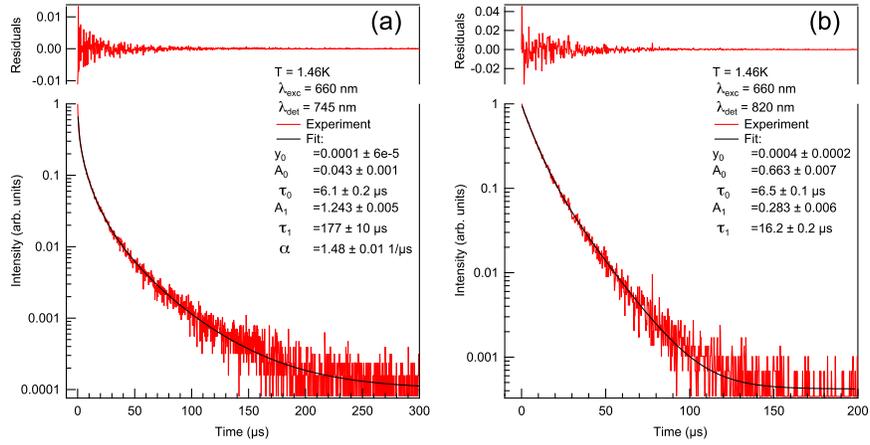}}
\caption{Decay kinetics recorded under excitation at 660\,nm. T\,=\,1.46\,K.
a) $\lambda{_{det}}$\,=\,745\,nm; a) $\lambda{_{det}}$\,=\,820\,nm.}
\end{figure}

We have shown above that under CW excitation at 660\,nm only two PL bands (P1 and P2) could be detected in the spectral range 670\,-\,900\,nm.
Nevertheless the assignment of the band P2 was somewhat ambiguous. For this reason we measured the decay kinetics of the PL bands P1 and P2 
with the use of the nanosecond OPO tuned to 660\,nm. The normalized data sets of these kinetics are shown in Fig\,10. It should be clear from the
comparison of Fig.\,9(b) and Fig.\,10(a) that the lifetime of P2 band is much shorter. Furthermore, the detailed analysis of the time dependence of its decay
have shown that the best fit function was  $I=y_{0} + A_0\exp(-t/\tau_{0}) + A_1\exp(-t/\tau_{1}-(\alpha t)^{3/8})$. The results of the fit to this function are
shown in Fig\,10(a) together with residuals. Most probably, the second term arises from the superposition of this band with P1 band (see Fig.\,3(b)). The last
term is similar to that for the decay kinetics of P3 band, but it takes account of the dipole-quadrupole nature of the donor-acceptor interaction. Finally, the
decay kinetics of P1 band can also be fit to the same function, which takes account of the dipole-quadrupole interaction. Nevertheless, the best fit was achieved to
the simple bi-exponential decay, probably, because of the low dynamic range of the data set. For this reason in Fig.\,10(a) we show this best fit to
the bi-exponential function together with its residuals.

\begin{figure}[htb]
\centerline{\includegraphics[width=12cm]{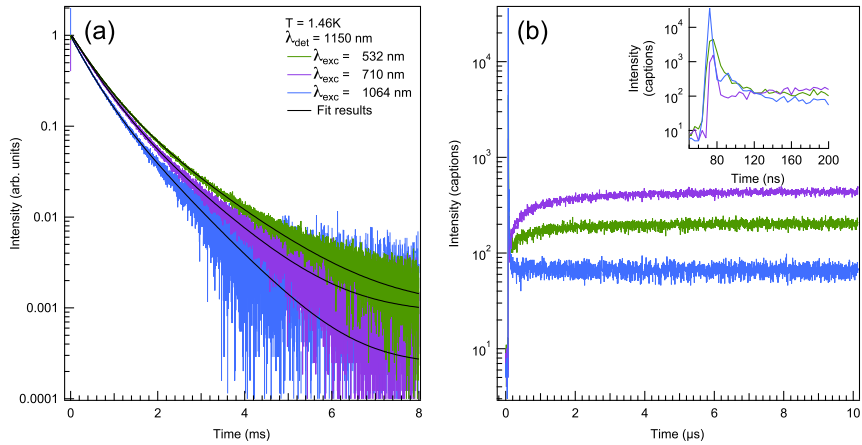}}
\caption{Decay kinetics recorded at 1150\,nm under excitation at 532, 710 and 1064\,nm. T\,=\,1.46\,K, excitation pulse width 8\,ns, pulse energy 100\,$\mu$J.
a) Time resolution 1.024\,$\mu$s; b) Time resolution 4\,ns, color coding corresponds to the left panel. Inset: expanded view of the fast initial stage.}
\end{figure}

In Fig.\,11 we show the decay kinetics measured at 1150\,nm at low temperature under excitation at 532, 710 and 1064\,nm recorded at two time resolutions:
1.024\,$\mu$s (Fig.\,11(a)) and 4\,ns.  The results of the numerical fit to the bi-exponential decay function for two temperatures (1.46 and 300\,K) are collected 
in Table\,2. Apart from the clearly pronounced bi-exponential decay, it is worth noting the following features: (i) the characteristic lifetimes are longer
at the shorter excitation WL's; (ii) the lifetimes are also longer at the high temperature; (iii) the PL rise time can be clearly detected at the excitation WL's 532
and 710\,nm, but not at 1064\,nm; (iv) the extremely fast and intense initial kinetics, as it is seen in the inset of Fig.\,11(b).
The three first features clearly indicate that there exist the upper states with the lifetime longer than the one of the emitting at 1150\,nm.
Furthermore, the longer lifetime at 300\,K under excitation at 1064\,nm also confirms the ET between two components of the band P0,
as it was shown in Fig.\,6. The fast initial kinetics were unresolved in our experiments and should be investigated by other methods
of the time resolved spectroscopy. At present we believe that the assignment of this fast initial kinetics to the \textit{hot luminescence} \cite{Rebane78} is most plausible. 

\begin {table}[h]
\caption {Summary of the bi-exponential fit parameters of the decay kinetics at 1150\,nm.} 
\label{Tab2} 
\begin{center}
\begin{tabular}{ |c|c|c|c|c|c|}
\hline
Excitation WL  & Temperature & A${_1}$  & $\tau{_1}$ & A${_2}$ & $\tau{_2}$  \\
(nm) &  (K) & (arb. units) & ($\mu$s) & (arb. units) & ($\mu$s) \\ \hline
\multirow{2}{*}{532} & 1.46 & 0.743 $\pm$ 0.002 & 498 $\pm$ 0.8 & 0.261 $\pm$ 0.002  & 1249 $\pm$  3   \\
&  300 & 0.747 $\pm$ 0.009 & 589 $\pm$ 1 & 0.260 $\pm$ 0.007 & 1381 $\pm$ 2   \\ \hline
\multirow{2}{*}{710} & 1.46 & 0.713 $\pm$ 0.004 & 472 $\pm$ 1 & 0.285 $\pm$ 0.004 & 1055 $\pm$ 6  \\
 & 300 & 0.715 $\pm$ 0.002  & 563 $\pm$ 1  & 0.281 $\pm$ 0.003 & 1207 $\pm$ 5   \\ \hline
 \multirow{2}{*}{1064} & 1.46 & 0.491 $\pm$ 0.033 & 366 $\pm$ 10 & 0.511 $\pm$ 0.033  & 794 $\pm$ 23 \\
  &  300 & 0.490 $\pm$ 0.159 & 544 $\pm$ 41 & 0.482 $\pm$ 0.16 & 945 $\pm$ 32 \\
\hline
\end{tabular}
\end{center}
\end {table}

\subsection{Energy level diagram}

To explain the described above experiments we propose an energy level diagram (ELD) shown in Fig.\,12. 
It consists of two interacting centers: Bi$^{+}$ ion and the defect, most probably associated to Bismuth doping.
On the left hand side we provide a scheme of the energy level splitting corresponding to Bi$^{+}$ ion.
Here the energy positions correspond to the PL bands observed in the experiment.
This diagram is very similar to the results of analysis of intra-configuration optical transitions in Pb$^0$(2) center in SrF$_2$ \cite{Bartram89}.
As in \cite{Bartram89} it is considered in the frame of the crystal field (CF) approximation, but it takes account of two important phenomena:
(i) the Jahn-Teller or pseudo Jahn-Teller effects (JTE and PJTE, respectively) \cite{Bersuker} and (ii) the ET between Bi$^{+}$ ion
and defects (ET$_{\textrm{da}}$). In Fig.\,12 for the sake of simplicity only two channels for the ET were shown, but it should be 
remembered that all excited states of Bi$^{+}$ ion are the subject of this ET. The relative efficiency of the ET from the particular
level can change the spectral shape of P0 band and its peak position, as it can be seen in Fig.\,1(a). 
The APES's corresponding to the energy levels at 737 and 1087\,nm (emission WL's) of Bi$^{+}$ ion are
shown in Fig.\,12 without the offset to simplify the diagram.
Note also that the definition of Bi$^{+}$ ion as a \textit{donor} is relative to the excitation wavelength, \textit{i.e.} when the system 
of such a donor-acceptor pair is excited in the range of 1070\,-\,1100\,nm the reverse transfer from defects to Bi$^{+}$ ions can take place 
resulting, for instance, in the low efficiency of the fiber laser at room temperature. The APES's drawn in the middle and right-most parts
should be regarded as a simple sketch, their exact parameters are unknown, here only the excitation and emission WL's are in the scale.  

\begin{figure}[htb]
\centerline{\includegraphics[width=10cm]{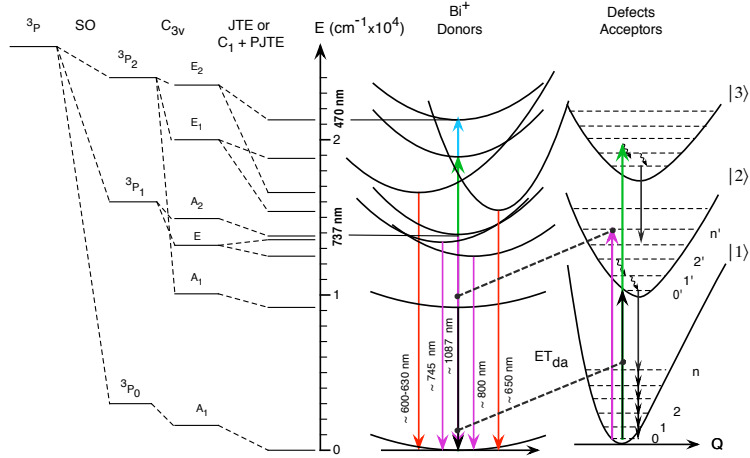}}
\caption{Energy level diagram of the donor-acceptor pair in Ga/Bi (Al/Bi) co-doped glass. SO - spin-orbit interaction; ET$_{\textrm{da}}$ - energy transfer between Bi$^+$ ions (donors), and defects (acceptors); 
JTE and PJTE are the Jahn-Teller and pseudo Jahn-Teller effects, respectively. }
\end{figure}

This simple CF approach can be justified in the studied case  by the interstitial character of the embedding of Bi-ions in the silica glass
network as it was discussed, for instance, in \cite{Sokolov08}. The latter is also supported by our technology process, which consists
in the low temperature sintering of the doped nano-porous silica xerogels. In the CF approximation the Bi-ion can be regarded as an ion which
does not form chemical bonds with the environment. 
This picture is not in contradiction with the model developed by E. Kustov et al. \cite{Kustov09,Kustov10}, where the chemical 
bonding of \textit{s}, \textit{p}, and \textit{d} electrons of Bismuth is taken into account explicitly in the frame of the semi-empirical
theory of molecular orbitals, but it can be considered as a rather simplified approach. 
Also, the low valence state of Bi-ions was suggested in \cite{Meng05, Peng11} on the basis of the phenomenological theory of the
optical basicity \cite{Duffy}, while the defects as a possible origin of NIR PL in glasses doped with different 6\textit{p} (Bi, Pb)
and 5\textit{p} (Sn, Sb) ions were suggested previously by Sharonov \textit{et al.}\cite{Sharonov08,Sharonov09}. 

In the frame of the CF approach one can find the similarity between the energy level diagrams in all types of the silica glass with a low content of Bismuth. 
Indeed, the general similarity was pointed out recently in \cite{Firstov11} and we can add here that the position of PL bands
and their relative intensity in Al/Bi co-doped silica glass are very close to that described in the present work.
Furthermore, the group of bands at 730\,-\,800\,nm in Al/Bi \cite{R11} and Ga/Bi  co-doped glasses
can be put in the correspondence to the bands at 830, 845 and 906\,nm (see Fig.\,2 and Fig.\,3 in \cite{R10}) 
in a silica sub-lattice.  We assign this group of PL bands to the ${^3}$P${_1}$ multiplet of Bi$^{+}$ ion split by the CF as it is shown
in Fig.\,12.  Note, that in the assumption of the axial local symmetry (C$_{3v}$, for instance) the CF results in the doublet, ${^3}$P${_1}$(E)
and singlet, ${^3}$P${_1}$(A${_2}$). Then the E state should be the subject of the more or less strong JTE
\cite{Bersuker}, resulting in a double well structure of APES. Obviously, in disordered hosts the local symmetry of  Bi$^{+}$ ion can be
lower than axial, but in this case there remains the possibility of the PJTE. This is also valid for all upper lying E states derived
from the ${^3}$P${_2}$ multiplet. The temperature dependence of the spectrum shown in Fig.\,2(b) suggests just this possibility.
The corresponding assignment of bands should be clear from Fig.\,12.

The analysis of the decay kinetics of the band P3 suggests its very long intrinsic lifetime, the longest among all measured in our
experiments. The transition ${^3}\textrm{P}{_1}(\textrm{A}{_2}) \rightarrow  {^3}\textrm{P}{_0} (\textrm{A}{_1})$ is rigorously forbidden
as it was discussed in details by Bartram \textit{et al.}\cite{Bartram89}, and for this reason its decay kinetics clearly
indicates the ET$_{\textrm{da}}$ to the defects. Note also, that the measured characteristic lifetimes of P0 band
under excitation at 1064\,nm are significantly shorter than that obtained under excitation at 532 and 710\,nm (see Table\,2). This justifies
the assignment of the lowest excited state of Bi$^{+}$ ion to A${_1}$ state derived from ${^3}$P${_2}$ multiplet. The transition
${^3}\textrm{P}{_2}(\textrm{A}{_1}) \rightarrow  {^3}\textrm{P}{_0} (\textrm{A}{_1})$ is allowed, but most probably has a
weak oscillator strength. 

The proposed ELD explains many other experiments, in particular, the low laser efficiency in alumino-silicate
fibers at room temperature, the strong temperature dependence of the laser efficiency and the unsaturable losses
\cite{Dvoyrin08,Dianov07}.  In the frame of the proposed model all these experimental results are the direct consequence  
of the ET$_{\textrm{da}}$ from defects to Bi$^{+}$ ion.
This should be clear from Fig.\,6. Furthermore, the experiments on optically detected magnetic resonance
(ODMR), albeit incorrectly interpreted in \cite{R09}, were performed just in this spectral range and also support
the assumption on the defects. Unfortunately, in the latter work the heavy doped samples were investigated and we
cannot exclude the contribution of clusters of Bismuth in the signal of ODMR.
  
Finally, we would like to emphasize that, strictly speaking, the proposed here ELD is valid for the glasses with a low content of Bismuth,
in other words for the \lq\lq lasing" glasses. The enhanced concentration of Bismuth  results  in the formation of dimers
and other nano-clusters of Bismuth. These clusters obviously have their own electronic structure and will contribute to the
spectrum of the absorption and PL that in turn leads to the quenching of the fiber laser efficiency, most probably also due to ET. 
Considering the pure silica glass doped with the Bismuth only, two scenarios are possible to explain the high efficiency
of the fiber lasers operating in the range of 1400\,-\,1450 nm (centers in a pure silica sub-lattice): 1) there are no defects associated
to Bismuth doping, then the fiber laser operates on the lowest transition of Bi$^{+}$ ion;
2) there is a large energy mismatch between corresponding absorption spectra of the defects and Bi$^{+}$ ion, so that the 
ET is inefficient. This point is not clarified up to now.

It should be also noted, that when the present work was completed, two articles on the PL in Bi-doped TlCl and CsI single
crystals appeared \cite{Plot13,Sokolov13}. Though the studied hosts are rather different from the silica glass, the conclusions
of the authors are similar to that in the present paper. Here we did not discuss the nature of the defect, which 
was considered in \cite{Plot13} as a complex of Bi$^+$ in Tl site and a negatively charged Cl vacancy in the nearest anion site. 
The energy transfer between Bi$^+$ ion and the defect, which was revealed in the present study suggests a different nature
of the latter, at least in a silica glass. Furthermore, the conclusions in \cite{Plot13,Sokolov13} were based on the quantum chemistry
calculations and they need to be verified experimentally, especially the point on the nature of the defect.

\section{Conclusion}
In conclusion, Ga/Bi co-doped silica glass was developed and investigated by continuous wave and time resolved photoluminescence spectroscopy in a wide temperature
range of 1.5\,-\,300\,K. The optical centers in this glass were identified as the analogues of the Al-connected centers in Al/Bi co-doped silica glass.
The investigations of the low temperature PL kinetics put in evidence the energy transfer between Bi$^+$ ions and optical centers
emitting in the range of 1000\,-\,1250\,nm (ET$_{\textrm{da}}$). The latter were identified as the defects associated to (or induced by)
the Bismuth doping. The consistent empiric model of Bi$^+$ ion and associated defect is put forward to explain
the peculiarities of the photoluminescence in a weakly doped with Bismuth silica glasses.

\section*{Acknowledgments}
I.R. is grateful to F. Real for the helpful discussions. The work was supported by  the \lq\lq Conseil R\'egional du Nord/Pas de Calais" and the \lq\lq Fonds Europ\'een de D\'eveloppement Economique des R\'egions" (FEDER) through the \lq\lq Contrat de Projets Etat R\'egion (CPER) 2007-2013" and by the \lq\lq Agence Nationale de la Recherche" through the contract ANR \lq\lq BOATS" 12BS04-0019-01.

\end{document}